\documentclass[aps,twocolumn,showpacs,floatfix]{revtex4}

\usepackage{graphicx}

\begin{document}
\title{Shock waves in a one-dimensional Bose gas: from a Bose-Einstein
condensate to a Tonks gas}
\author{Bogdan Damski}
\affiliation{
Theory Division, Los Alamos National Laboratory, MS-B213, Los Alamos, NM 87545, USA
}
\begin{abstract}
We derive and analyze shock-wave solutions of  hydrodynamic equations 
describing repulsively interacting one dimensional  Bose gas. 
We  also use the number-conserving Bogolubov approach to 
verify  accuracy of the Gross-Pitaevskii equation in shock wave problems. We show that 
quantum corrections to dynamics of  shocks (dark-shock-originated solitons) 
in a Bose-Einstein condensate  are negligible (important) for a realistic set 
of system parameters.
We point out possible signatures of a Bose-Einstein condensate -- Tonks 
crossover in shock dynamics. 
Our findings can be directly verified in different experimental setups.

\end{abstract}
\pacs{03.75.Kk,47.35.+i,43.35.+d}
\maketitle

\section{Introduction}
\label{sec1}
Physics of one dimensional (1D) Bose gases attracts more and more attention due
to both  very interesting phenomena that appear in these systems \cite{korepin}, 
and a very recent experimental progress toward realization of Tonks gases
\cite{esslinger,paredes,raizen}.
In this paper we provide a unified description of basic shock waves properties  
in  1D Bose gases interacting with arbitrary strength.
Previous works on this subject were limited to either  weakly interacting
gases \cite{bodzio_b,gammal,konotop,menotti,el} 
or  the strongly interacting  ones \cite{bodzio_f}.
We also describe influence of depleted atoms  on
dynamics of a Bose-Einstein condensate (BEC) shocks. Finally, we derive a simple 
expression for  speed of propagation of arbitrarily shaped density
pulses.

Sec. \ref{sec2} describes
theoretical basics of an approximate  hydrodynamic  approach. 
Sec. \ref{sec3} discusses
shock-wave solutions of this approach  on  specific 
examples. Sec. \ref{sec4} describes propagation  of arbitrarily
shaped pulses. Sec. \ref{sec5} explains experimental methods used 
for generation of shock structures, while Sec. \ref{sec6} presents
quantum corrections to   shock 
solutions of the  Gross-Pitaevskii   equation.
Finally, Sec. \ref{sec7} provides a summary of this paper.

\section{The model}
\label{sec2}
We consider delta interacting Bose gas 
in a  1D box. Its  experimental realization 
is possible due to recent experimental progress
in trapping of small samples of ultracold  bosonic atoms  
in a box-like optical trap \cite{raizen}.
Another exciting experimental setup 
may come from a paper of Gupta {\it et al} \cite{gupta}, where  successful 
realization of a ring shaped magnetic trap filled with a BEC
was recently reported.

\begin{figure}
\includegraphics[angle=-0,scale=0.35, clip=true]{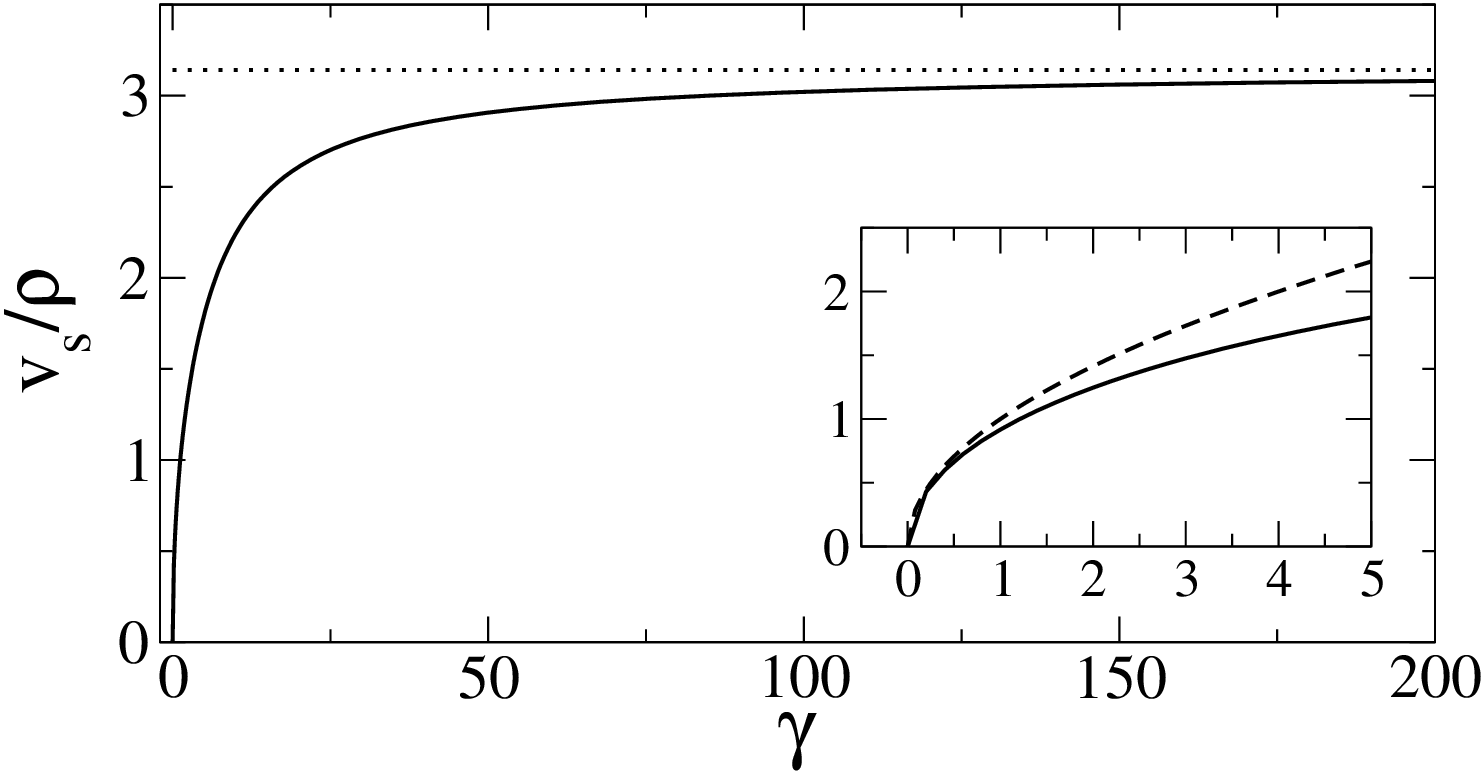}
\caption{Sound velocity divided by density. In the BEC limit
($\gamma\to0$) it is $\sqrt{\gamma}$, while in the Tonks limit
($\gamma\to+\infty$) it equals $\pi$.
Solid line: (\ref{exact}), dots:  Tonks value, dashed line:  $\sqrt{\gamma}$.
For units see Appendix \ref{a2}.
}
\label{fig1}
\end{figure}

The Hamiltonian of our system in dimensionless quantities 
(see Appendix \ref{a2} for units)
reads: 
\begin{equation}
\hat{H}= -\frac{1}{2}\sum_{i=1}^N\frac{\partial^2}{\partial x_i^2}
+a \sum_{i<j} \delta(x_i-x_j),
\label{manybody}
\end{equation}
where $a>0$ is the interaction coupling \cite{olshanii}. 
It turns out that there is only one parameter that determines system
static properties:
\begin{equation}
\gamma= \frac{a}{\rho},
\end{equation}
where $\rho$ is atomic density 
normalized to $N\gg1$. In the limit of $\gamma\to0$
our system is a Bose-Einstein condensate, while when $\gamma\to+\infty$
it is a Tonks gas, whose properties are captured 
by the Fermi-Bose mapping theorem (FBMT) \cite{bfmt}. To see how big is a BEC-Tonks
crossover one can look at  sound velocity and compare it to 
BEC and Tonks predictions. Exact expression for sound velocity  
\cite{lieb} (see also Appendix \ref{a2}) is 
\begin{equation}
v_s= \rho \sqrt{3 e(\gamma) - 2\gamma\frac{de}{d\gamma} 
+\frac{1}{2}\gamma^2\frac{d^2e}{d\gamma^2}},
\label{exact}
\end{equation}
where
$e(\gamma)$ is defined by Lieb and Liniger in \cite{liebliniger}. 
The $v_s/\rho$ is depicted in Fig.
\ref{fig1} -- notice how slowly the system enters a Tonks regime.
A large BEC-Tonks crossover, $\gamma\sim(1,50)$, provides 
us motivation for studies of shock waves outside  BEC
\cite{bodzio_b,gammal,konotop,menotti,el} and Tonks \cite{bodzio_f} limits.

Many-body
solutions of (\ref{manybody}) at zero absolute temperature
have been analyzed in  classic papers \cite{liebliniger,lieb}. 
These solutions are based on the Bethe ansatz and 
allow for analytical extraction of different static system properties.
Unfortunately, they are too complicated for description of system dynamics. 

To simplify the problem a proper hydrodynamical approach can be worked out 
 \cite{olshani2,kohn}
and leads to the following set of equations
\begin{equation}
\frac{\partial \rho}{\partial t} + \frac{\partial}{\partial x}\left(v\rho\right)
 =0,
 \label{cont}
\end{equation}
\begin{equation}
\label{euler}
 \frac{\partial v}{\partial t} + \frac{\partial}{\partial x}\left(\frac{1}{2} 
 v^2\right)+\frac{\partial}{\partial x} \left(\mu(\rho)+V_l
 -\frac{1}{2}\frac{\partial_x^2 \sqrt{\rho}}{\sqrt{\rho}}\right)=0
\end{equation}
where 
\begin{equation}
\label{muuu}
\mu(\rho)=\frac{1}{2}\rho^2\left(3e(\gamma)-\gamma\frac{de}{d\gamma}\right),
\end{equation}
and $V_l$ is an external potential acting on atoms.
A nice property of these equations is that they exactly reproduce 
sound velocity (\ref{exact}), while the problems with 
their usage for description of shock propagation (not formation)
come from their derivation  valid in
the long-wavelength limit, where the quantum pressure (QP) term,
\begin{equation}
\label{qpressure}
\frac{1}{2}\frac{\partial_x^2 \sqrt{\rho}}{\sqrt{\rho}},
\end{equation}
is negligible. 

To get more insight into  physics described by (\ref{euler}) and
validity of a QP  term we 
 look at  BEC and Tonks limits. In the BEC case, 
$\mu(\rho)\to a\rho$ and Eqs. (\ref{cont},\ref{euler}) can be obtained 
by time-dependent variational principle applied to a product 
 wave function
\begin{equation}
\label{spwf}
\Psi(x_1,\dots,x_N,t)=\sqrt{N}\phi(x_1,t)\cdots\phi(x_N,t),
\end{equation}
$\int{\rm d}x\,|\phi(x,t)|^2=1$. It results in time-dependent Gross-Pitaevskii
equation (\ref{gp}).
Then  substitution of $\phi=\sqrt{\rho}\exp(i\chi)$ and $v=\partial_x\chi$
into (\ref{gp}) gives  a 
BEC version of (\ref{cont},\ref{euler}). In other terms, $\rho$ 
in (\ref{cont},\ref{euler}) is a single particle
density defined in a many-body theory as 
\begin{equation}
\label{spd}
\rho(x,t)= \int {\rm d}x_2\cdots{\rm
d}x_N|\Psi(x,x_2,\dots,x_N,t)|^2,
\end{equation}
which equals $N|\phi(x,t)|^2$ in the BEC limit. 
A quantum pressure term in this limit is rigorously 
derived \cite{dalfovo}.

In the Tonks regime one has $\mu(\rho)\to \rho^2\pi^2/2$, which 
was first found by renormalization 
group approach in \cite{kolomeisky,kolomeisky1}, and then used in a number of  
papers, e.g.,  \cite{bodzio_f,proukakis,tosi}. 
Now derivation of 
(\ref{euler}), do not involve any sort of product 
simplification (\ref{spwf}) of a wave-function. 
Indeed, a Fermi-Bose mapping
theorem \cite{bfmt} implies, e.g., that if
$\Psi_F(x_1,\dots,x_N)$ is a ground state 
wave-function of noninteracting fermions placed in the same potential as a Tonks gas, then
a ground state Tonks wave-function is $|\Psi_F(x_1,\dots,x_N)|$.
Therefore, even a simple description of a Tonks gas based on the single 
particle density (\ref{spd})  involves 
$N$ orthogonal single particle orbitals instead of a single one, $\phi(x)$,
used in the BEC case.

In  Tonks limit the QP term was shown to lead to unphysical
density oscillations \cite{inter,bodzio_f}.
From  knowledge that the quantum pressure term is present in the BEC
limit $\gamma\to0$ and absent in Tonks regime $\gamma\to+\infty$
it is clear that Eqs. (\ref{cont},\ref{euler}) can not describe 
shock propagation for arbitrary  $\gamma$. Nonetheless, shock formation 
from density perturbations that are initially wide can be 
successfully done, and will be described below. In this case the
QP term is unimportant roughly up to shock formation.
Due to lack of theoretical concepts for getting exact
time-dependent solutions for the system of interest, we consider future  experimental 
results as the best verification of our calculations based on
hydrodynamic equations.

\section{Shock wave solutions}
\label{sec3}
The quantum pressure term (\ref{qpressure}) is important only when 
density changes occur on length scales smaller then the characteristic length
 given by 
\begin{equation}
\label{healing}
\xi(\rho)= \frac{1}{\sqrt{2 \mu(\rho)}}.
\end{equation}
In a BEC limit $\xi(\rho)$ is called a healing length,
and we propose to use the same name regardless of $\gamma$. 

Now we want to solve Eqs. (\ref{cont},\ref{euler},\ref{muuu}).  
Assuming that the perturbation under consideration is, at least initially,
broad compared to $\xi$, the quantum pressure term in (\ref{euler}) 
can be neglected. Following standard methods \cite{landau} one gets
\begin{equation}
\label{nonlin}
\rho(x,t)= f\left(x-\left(
\int d\rho \sqrt{\frac{1}{\rho}\frac{\partial \mu}{\partial \rho}}
+\sqrt{\rho\frac{\partial \mu}{\partial \rho}}\right)t\right),
\end{equation}
\begin{equation}
\label{velocity}
v(x,t) = \int d\rho \sqrt{\frac{1}{\rho}\frac{\partial \mu}{\partial \rho}}.
\end{equation}

To proceed further with analytical calculations we  use the following 
relations:
\begin{equation}
\label{bec}
0<\gamma<\gamma_c:  \ \ 
\sqrt{\frac{1}{\rho}\frac{\partial \mu}{\partial\rho}} \approx
\sqrt{\gamma} - \frac{\gamma}{4\pi} -\frac{\gamma^{3/2}}{32\pi^2}
-\frac{\gamma^2}{128\pi^3}, 
\end{equation}
\begin{equation}
\label{tonks}
\gamma_c\le\gamma:  \ \
\sqrt{\frac{1}{\rho}\frac{\partial \mu}{\partial\rho}} \approx
\frac{\pi}{(1+2/\gamma)^2},
\end{equation}
where $\gamma_c\approx14.5$ from the requirement that both expansions match 
at $\gamma_c$. Expression (\ref{bec}) is extracted from 
Lieb's observation \cite{lieb} that for $\gamma< \sim10$
\begin{equation}
\label{obs}
\sqrt{\frac{1}{\rho}\partial_\rho\mu}\cong 
\sqrt{\gamma-\gamma^{3/2}/(2\pi)}.
\end{equation}
Since (\ref{obs}) 
complicates further calculations, we expanded it into a series
around $\sqrt{\gamma}/(2\pi)$ equal to zero. 
Expression (\ref{tonks}) is taken directly from Lieb and Liniger paper
\cite{liebliniger} -- see Eqs. (3.32) and Appendix \ref{a2}. 

To test above approximations we compare 
the shock wave solutions that neglect the QP term 
and use approximate expression for 
$\sqrt{\frac{1}{\rho}\frac{\partial \mu}{\partial\rho}}$, 
to the full numerical solution of 
hydrodynamic equations (\ref{cont},\ref{euler},\ref{muuu}).
Initially we determine velocity field from (\ref{velocity}) and choose 
\begin{equation}
\label{hyperbolicus}
\rho(x,0)=\rho_0+\frac{\eta\rho_0}{\cosh(x/\sigma)^2} \ \ , \ \ \eta>0,
\end{equation}
with $\sigma\gg\xi(\rho)$ (calculations for $\eta<0$ can
be easily repeated).
The background density $\rho_0$ is found from
the normalization condition: $\int_{-l}^ldx\,\rho(x,0)=N$, where we have assumed
that periodic box has boundaries at $\pm l$. For well
localized perturbations being of interest from now on one gets 
$\rho_0 = \frac{N}{2 (l + \eta \sigma)}$. 
Finally, it is  convenient to
define  relative density
$$
\varrho=\frac{\rho}{\rho_0}.
$$
Now we  consider separately $\gamma>\gamma_c$ and $\gamma<\gamma_c$ cases.

\subsection{$\gamma\ge\gamma_c$ case}
\label{3a}

\begin{figure}
\includegraphics[angle=-0,scale=0.35, clip=true]{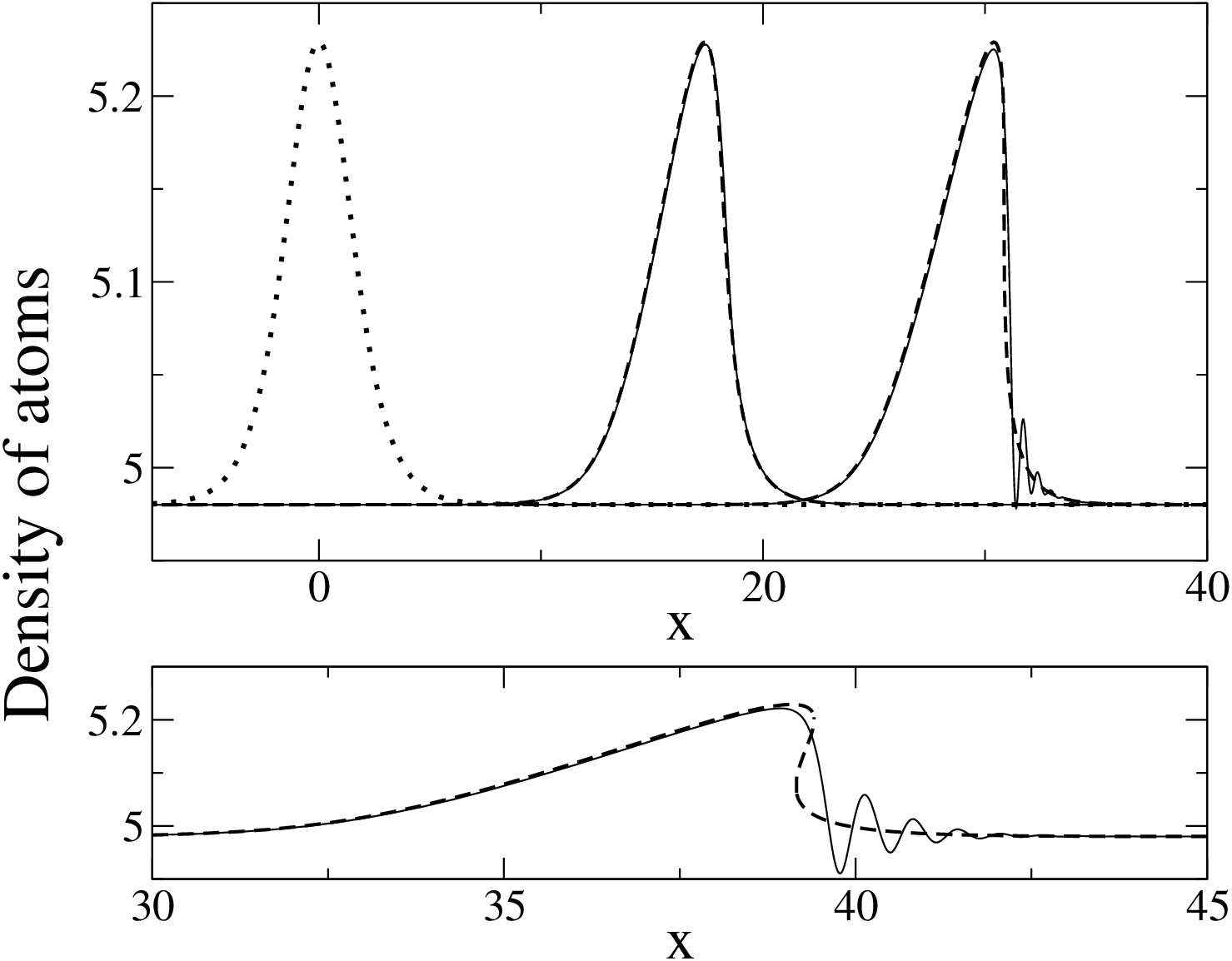}
\caption{Dotted line: initial density profile (\ref{hyperbolicus}). Solid line:
solution of Eqs. (\ref{cont},\ref{euler}) with density 
at $t=0$ given by (\ref{hyperbolicus}) and velocity field determined from 
(\ref{velocity}). Dashed line: implicit solution (\ref{hypimpl})
multiplied by $\rho_0$. 
Upper plot: subsequent density profiles 
correspond to $t=0, 1.155, 2.02$. Lower plot: $t=2.6$. 
Time of shock creation equals $t_s=2.03$, 
$N=499$, $\gamma_0=30$, $\sigma= 2$, $l=50$, $\eta=0.05$.
For units see Appendix \ref{a2}.
}
\label{fig2}
\end{figure}

The healing length is found from 
(\ref{healing}) and  Eqs. (3.32) of \cite{liebliniger} subjected to 
rescalings of Appendix \ref{a2}  
\begin{equation}
\label{healing_tonks}
\xi(\rho)= \frac{1}{\pi\rho}\left(1+ \frac{8}{3\gamma} + \frac{2}{3\gamma^2}\right),
\end{equation}
where the first term corresponds to the Fermi length of a non-interacting
Fermi gas -- a result that may be  expected from the FBMT \cite{bfmt}. 

\noindent Rewriting implicit solution (\ref{nonlin}) to the form
$$
\varrho(x,t)= f\left(x-c_\infty w_\infty(\varrho)\,t\right),
$$
where $c_\infty=\pi\rho_0$ is a background sound velocity of a Tonks gas,
one gets from (\ref{nonlin}) and (\ref{tonks}) 
$$
w_\infty(\varrho)= \frac{2\varrho+s\varrho^2-1}{(1+s\varrho)^2(1+s)} \ \ \ , \ \ \
s=\frac{2}{\gamma_0}.
$$

Taking (\ref{hyperbolicus}) as an initial density profile, the implicit
shock-wave solution reads:
\begin{equation}
\label{hypimpl}
\varrho(x,t)= 1+ \frac{\eta}{\cosh[(x-c_\infty w_\infty(\varrho)t)/\sigma]^2}.
\end{equation}

Although, the explicit form of $\varrho(x,t)$ can not be extracted analytically, 
important properties of pulse dynamics can be  determined.
First of all, impulse amplitude, $1+\eta$, is constant in time.
Second, speed of impulse maximum for any $\eta>0$ equals
$$
{\cal V}(\eta)= c_\infty w_\infty(1+\eta),
$$
and applicability of this expression is not limited to 
$1/\cosh(x/\sigma)^2$ perturbations only -- Sec. \ref{sec4}. 
Third, the width of the impulse at  given density is constant during
propagation- a property that we missed in earlier papers
\cite{bodzio_b,bodzio_f}. 
Fourth,  impulse tails propagate roughly with the
background sound velocity equal to $c_\infty w_\infty(1)$, while the impulse
maximum moves with the speed $c_\infty w_\infty(1+\eta)$. Since $w(\varrho)$
monotonically increases with $\varrho$, the impulse deforms its shape  so that
a shock wave front forms, i. e., $|\partial_x\rho(x,t)|=+\infty$ at one point.

Time and position of shock-wave creation can be extracted from
equations \cite{landau}:
\begin{equation}
\label{shock}
\partial_\rho x(\rho)= 0 \ \ \ , \ \ \ \partial^2_\rho x(\rho)= 0.
\end{equation}
Their  solution gives:
 density $\rho_s$ at which density profile becomes locally 
vertical, and time $t_s$ at which this occurs:
\begin{equation}
\varrho_s= \frac{3+4s\eta+9s-\sqrt{9(1+s)^2+16s^2\eta^2}}{6s},
\end{equation}
\begin{equation}
t_s= \frac{\sigma\sqrt{\eta}}{4 c_\infty}\frac{(1+s\varrho_s)^3}{(\varrho_s-1)
\sqrt{1+\eta-\varrho_s}}.
\end{equation}
These expressions close to a Tonks gas limit take a simple form:
$$
\rho_s= 1+ \frac{2}{3}\eta + {\cal O}\left(\frac{1}{\gamma}\right) \ \ , \ \
t_s= \frac{3\sqrt{3}}{8}\frac{\sigma}{\eta c_\infty} + {\cal
O}\left(\frac{1}{\gamma}\right).
$$

Comparison of analytical solution (\ref{hypimpl}) 
and full numerical one
based on  hydrodynamic equations (\ref{cont},\ref{euler},\ref{muuu}) 
is presented in Fig. \ref{fig2}. As easily seen, there is a good
agreement between hydrodynamical  solution and a shock-wave 
one until the moment of shock creation. Then discrepancy 
increases due to appearance of density oscillations triggered by the 
QP  term (\ref{qpressure}) neglected in derivation 
of (\ref{hypimpl}). 
Since presence of the QP term for large $\gamma$, e.g., $\gamma\ge\gamma_c$,
is questionable, it is an interesting open question whether 
density oscillations in the form presented in Fig. \ref{fig2}
survive or not for any $\gamma\ge\gamma_c$  considered here.

\subsection{$\gamma<\gamma_c$ case}
\label{sec3b}
The healing length reads 
\begin{equation}
\label{healing_bec}
\xi(\rho)= 
\frac{1}{\sqrt{2a\rho}}\left(1+  \frac{\sqrt{\gamma}}{2\pi}+\frac{3\gamma}{8\pi^2}\right),
\end{equation}
derived by  extraction of $\mu(\rho)$ from (\ref{obs})
and subsequent Taylor expansion of (\ref{healing}). 
Naturally, the first term in (\ref{healing_bec}) corresponds to the BEC  
healing length. 

The implicit shock-wave solution has the form:
\begin{equation}
\label{i}
\varrho(x,t)= f\left(x- c_0w_0(\varrho)t\right),
\end{equation}
where $c_0=\sqrt{a\rho_0}$ is the background speed of sound 
in the limit of $\gamma\to0$. Combining (\ref{nonlin}) and 
(\ref{bec}) one gets 
\begin{eqnarray}
w_0(\varrho)=&& 3\sqrt{\varrho}-2-
\frac{\sqrt{\gamma_0}}{4\pi}\left(\ln\varrho+1\right)+
\frac{\gamma_0}{32\pi^2}\left(\sqrt{\frac{1}{\varrho}}-2\right)
\nonumber \\ &&-\frac{\gamma_0^{3/2}}{128\pi^3}.
\end{eqnarray}
To verify  accuracy of (\ref{i}) 
we have simulated dynamics for the system with $\gamma_0=a/\rho_0=5$,
i. e., for $\gamma_0$ outside the BEC mean-field regime (see inset of
Fig. \ref{fig1}), but small enough
to stay clearly within  $\gamma<\gamma_c$ parameter range. 
Agreement between full numerical solution of Eqs.
(\ref{cont},\ref{euler}) and implicit shock solution is 
 good   before shock formation. Close to  time
of shock creation discrepancies show up in the oscillatory 
region missed in (\ref{i}). Qualitatively, the plot 
that presents these results is the same as Fig. \ref{fig2}.

Since  dynamics of initial density profile is qualitatively the
same as in
the $\gamma\ge\gamma_c$ case, we notice  that  impulse maximum 
moves now for any $\eta>0$ with velocity 
$$
{\cal V}(\eta)= c_0 w_0(1+\eta).
$$

The explicit solution of shock equations (\ref{shock}) can be  
found in the limit of $\gamma_0\to0$, but still arbitrary $\eta>0$:
$$
\rho_s=\frac{1+\eta+\sqrt{(1+\eta)(9+\eta)}}{4},
$$
$$
t_s= \frac{\sigma\sqrt{\eta\rho_s}}{3c_0(\rho_s-1)\sqrt{1+\eta-\rho_s}}.
$$
Interestingly, for a gaussian like initial  impulse
discussed in \cite{bodzio_b}  explicit expressions
for $\rho_s$ and $t_s$ in this limit were beyond the reach. 
Finally, we note  that the QP
term for $\gamma\to0$ is certainly present, but it is hard to say
whether it survives in (\ref{euler}) for 
any $\gamma<\gamma_c$. Since the QP term affects shock dynamics 
strongly, an experiment should clarify this uncertainty.
\section{Propagation of pulses of arbitrary shape}
\label{sec4}
In this section we describe some general properties of 
shock-wave solutions (\ref{nonlin},\ref{velocity}). To this aim  we consider
density profiles satisfying 
\begin{equation}
\label{impl}
\rho(x,t)= f(x- W(\rho)\,t)=f(\zeta).
\end{equation}
Suppose that there are $n$  extrema (minima, maxima),
 placed on a background density $\rho_0$,
 in the initial  density profile $\rho(x,0)$. 
 Let us denote positions of these extrema  at $t=0$ as $x_i(0)$ and 
 set  $\rho_i=\rho(x_i(0),0)$.
 It means that $\partial_x f(x)|_{x_i(0)}=0$ and 
 $\partial^2_x f(x)|_{x_i(0)}\neq0$.
 At $t>0$ one gets from (\ref{impl})
$$\partial_x \rho(x, t)= \frac{\frac{\partial f(\zeta)}{\partial \zeta}}
{1+\partial_\rho W(\rho) \frac{\partial
f(\zeta)}{\partial \zeta} t   },$$
which obviously equals zero iff  $\zeta=x_i(0)$. 
It implies that
$x_i(t)= x_i(0)+ W(\rho_i)\, t$. 
Furthermore, we have $\partial^2_x \rho(x,t)|_{x_i(t)}=\partial^2_x \rho(x,0)|_{x_i(0)}\neq0$, 
which proves that 
extrema do not change into saddle points in the course of time evolution.
This means that 
velocity of impulse extrema equals $W(\rho_i)$, where
$\rho_i=\rho(x_i(t),t)=\rho(x_i(0),0)$.

Since $W(\rho)$ is shape independent, 
we conclude that for any $\gamma$ density extrema propagate 
with constant amplitude and speed equal to $W(\rho_i)$. This speed is 
well approximated as 
\begin{equation}
\label{bec_speed}
0<\gamma<\gamma_c:  \ \ \ c_0 w_0(\rho_i/\rho_0),
\end{equation}
\begin{equation}
\label{tonks_speed}
\gamma>\gamma_c:  \ \ \ c_\infty w_\infty(\rho_i/\rho_0),
\end{equation}
for $\gamma$'s a little off  $\gamma_c$.

Finally, it is important to stress that above statements
concerning the amplitude and velocity of density extrema 
are valid at least  as long
as the implicit solution works, i. e., up to the time of shock creation.
After that time, they are valid in  regions far enough from shock
structures. 

\section{Experimental realization}
\label{sec5}
As discussed in \cite{bodzio_b,bodzio_f}, 
experimental creation of 
matter-wave packets undergoing  shock-wave dynamics is  straightforward.
The idea is to cool  atoms in an additional external potential created by a 
well-detuned laser beam, and then suddenly turn the beam off. The term
suddenly, means on  time scale much smaller than  time of sound propagation 
through the perturbation.

The initial density of atoms resulting from an external laser potential $V_l$ is 
determined as follows. Combining density $\rho$ and velocity field
$v=\partial_x\chi$ into $\phi=\sqrt{\rho}\exp(i\chi)$ the 
Eqs. (\ref{cont},\ref{euler}) can be rewritten to the standard form
\begin{equation}
i\partial_t\phi=-\frac{1}{2}\partial_x^2\phi+ V_l(x) \phi+ \mu(\rho)\phi,
\label{gp_general}
\end{equation}
which time-independent version reads 
$$
-\frac{1}{2}\partial_x^2\phi+ V_l(x)\phi+ \mu(\rho)\phi= \tilde{\mu}\phi,
$$
with $\tilde{\mu}$ being a chemical potential. Assuming that the laser potential 
induces density changes on  length scales larger than the healing length
(\ref{healing}), one determines an initial density profile from 
$$
V_l(x)+ \mu(\rho)= \tilde{\mu},
$$
with $\tilde{\mu}$ found from  $\int dx\, \rho=N$.
Naturally, velocity field equals zero in an initial state considered here.

A straightforward generalization of  results from \cite{bodzio_b,bodzio_f}
shows that if the initial density profile is $\rho(x,0)=\rho_0+h(x)$ with 
${\rm max}|h(x)|\ll\rho_0$ then 
$\rho(x,t)=\rho_0+ h\left(x-v_s(\rho_0) t\right)/2+h\left(x+v_s(\rho_0) t\right)/2$, 
with $v_s(\rho_0)$ 
being  background sound velocity (\ref{exact}). 
Therefore, there are  left and right moving perturbations. Due to  mirror
symmetry, we
can forget about the left moving one  once the pulses are well separated. 
The right-moving pulse  is described by $\rho_0+ h(x-v_s(\rho_0) t)/2$ as long as
$t\ll t_s$, since even arbitrarily small density perturbations
experience a shock deformation after long enough evolution.

In  BEC and Tonks limits
qualitatively the same splitting process  takes place even when 
${\rm max}|h(x)|/\rho_0\sim1$ \cite{bodzio_b,bodzio_f}.
When the initial density profile is $\rho(x,0)=\rho_0+h(x)$, the 
right moving pulse is described by 
$\rho(x,t)\approx \rho_0+f(x-W(\rho)t)$ with $W(\rho)$ being the same as in (\ref{impl})
and $f(x)=h(x)/2$. 
We checked 
by solving hydrodynamic equations (\ref{cont},\ref{euler})  that 
above described  splitting process 
works qualitatively the same way for any $\gamma$, as it works for
$\gamma\to0,+\infty$ \cite{bodzio_b,bodzio_f}.
This observation explains  how  we imagine practical generation of matter-wave pulses
described in previous sections.

\section{Applicability of the Gross-Pitaevskii mean-field approach 
         to shock problems}
\label{sec6}

In the BEC limit system dynamics in the mean-field approximation (\ref{spwf}) is
described by the following version  of Eq. (\ref{gp_general})
\begin{equation}
i\partial_t\phi=-\frac{1}{2}\partial_x^2\phi+ V_l(x) \phi+  aN|\phi|^2\phi\equiv H_{GP}\phi,
\label{gp}
\end{equation}
where $\int{\rm d}x\,|\phi(x,t)|^2=1$.
This is the Gross-Pitaevskii (GP) equation, which can be rigorously 
derived by different means -- see, e.g.,  \cite{dalfovo,castin}.

To go beyond the mean-field approximation we employ a second quantization
formalism, i. e., we transform  Hamiltonian (\ref{manybody}) with an 
additional
potential term, $V_l(x)$, into the form
\begin{equation}
\label{sq}
\hat{H}= \int dx \, \hat{\Psi}^\dag\left(-\frac{1}{2}\frac{\partial^2}{\partial x^2}+
V_l(x)+ 
\frac{a}{2}\hat{\Psi}^\dag\hat{\Psi}\right)\hat{\Psi},
\end{equation}
where $\hat{\Psi}$ is a field operator. In the number conserving
Bogolubov approach \cite{castin} it reads
$$
\hat{\Psi}= \hat{a}_0 \phi + \delta\hat{\Psi},
$$
with $\hat{a}_0$ annihilating a particle from a condensate, and $\delta\hat{\Psi}$
annihilating a particle from  modes orthogonal to the condensate one.
Naturally, $\phi$ is a condensate mode, i. e., it is an eigenstate of a single
particle density matrix, $\langle \hat{\Psi}^\dag(x')\hat{\Psi}(x)\rangle$,
to the highest eigenvalue ($\lambda_0\sim N$):
$$
\int {\rm d}x' \, \langle \hat{\Psi}^\dag(x')\hat{\Psi}(x)\rangle \phi(x')=
\lambda_0\phi(x).
$$
The single particle density of atoms, $\langle \hat{\Psi}^\dag(x)\hat{\Psi}(x)\rangle$,
equals
\begin{equation}
\label{ddensity0}
\langle \hat{a}_0^\dag\hat{a}_0\rangle 
|\phi(x)|^2 + \sum_k |v_k(x)|^2,
\end{equation}
where the first term accounts for condensate density while the second one 
corresponds to  density of depleted atoms. In further calculations we need
both $u_k$ and $v_k$, e. g., 
for finite temperature calculations. Indeed, 
density of atoms at sufficiently low temperature $T$ ($\lambda_0(T)/N \sim1$)
equals \cite{castin}
\begin{equation}
\label{ddensityT}
\langle \hat{a}_0^\dag\hat{a}_0\rangle 
|\phi(x)|^2+  \sum_k  \langle \hat{b}_k^\dag \hat{b}_k\rangle  \langle u_k|u_k\rangle+
\sum_k \langle \hat{b}_k^\dag \hat{b}_k+1\rangle \langle v_k|v_k\rangle,
\end{equation}
where $\langle \hat{b}_k^\dag \hat{b}_k\rangle=[\exp(\omega_k/(k_BT)-1]^{-1}$.
To get $u_k$, $v_k$ and $\omega_k$ 
one first finds a condensate mode from the 
time-independent GP equation, $H_{GP}\phi=\tilde{\mu}\phi$, and then constructs the matrix 
$$
{\cal L}=\left(
\begin{array}{cc}
H_{GP}+ aN Q|\phi|^2-\tilde{\mu} & aN Q\phi^2 \\
-aN Q^*\phi^{*2} & -H_{GP}- aN Q^*|\phi|^2+\tilde{\mu}
\end{array}
\right), 
$$
where  $Q$ is a projector to  space orthogonal to a condensate mode:
$Q\psi= \psi-\phi\langle\phi|\psi\rangle$. The modes and frequencies
are found from the eigen equation
$$
{\cal L}
\left( 
\begin{array}{c}
u_k \\ v_k
\end{array}
\right)= 
\omega_m \left( 
\begin{array}{c}
u_k \\ v_k
\end{array}
\right),
$$
while their dynamics is captured by 
$$
i\frac{\partial}{\partial t} 
\left( 
\begin{array}{c}
u_k \\ v_k
\end{array}
\right)= 
{\cal L}
\left( 
\begin{array}{c}
u_k \\ v_k
\end{array}
\right),
$$
where $\tilde{\mu}$ is set to zero in ${\cal L}$,
and the calculation of operator ${\cal L}$ at every time step requires 
simultaneous solution of the time-dependent Gross-Pitaevskii equation.
Expressions for densities of atoms (\ref{ddensity0}) and (\ref{ddensityT})
are unchanged in a time dependent case.
For a list of $u$, $v$ modes properties, and 
details of the number-conserving Bogolubov approach see 
\cite{castin}.

It is important to realize what are limitations of the
Bogolubov approach. To simplify notation we  discuss 
$T=0$ case now (the results for $T\neq0$ can be easily obtained).
The Bogolubov method fails in the following two situations.
First, the global breakdown, shows up when  total number 
of depleted atoms, i. e., $\int dx \sum_k |v_k(x)|^2$,
becomes comparable to   number of atoms in the system.
This limitation comes directly from derivation 
of  Bogolubov Hamiltonian \cite{castin}.
Second, local breakdown, happens when density of depleted atoms 
becomes comparable to  condensate density on a length scale of 
the order of the healing length:
\begin{equation}
\label{instability}
\int_{x-f\xi}^{x+f\xi} dx' N|\phi(x')|^2\sim \int_{x-f\xi}^{x+f\xi} dx' \sum_k |v_k(x')|^2,
\end{equation}
with $f={\cal O}(1)$. To clarify (\ref{instability}), we 
notice that in the Bogolubov approach an  impact of noncondensed
atoms on condensed ones is neglected. Indeed, to get Bogolubov dynamics 
one solves a time-dependent GP equation and then puts this solution 
into equations of motion of noncondensed atoms. In this way  noncondensed
particles are influenced by  condensed ones but not other way round. 
Next order corrections take into account change of 
condensate density as a result of 
repulsive interactions between condensed and noncondensed clouds, and 
lead to modifications of the GP equation. The correction to the right hand
side of (\ref{gp}) is qualitatively of the form 
$V_{\rm depl} \phi$ (Eqs. (95,96) in \cite{castin}), where
\begin{equation}
\label{vdepl}
V_{\rm depl}(x)\cong 2 a\sum_k |v_k(x)|^2,
\end{equation}
so that  $V_{\rm depl}(x)$ can be regarded as 
an external potential acting on a condensate. Extension of
changes of  condensate density induced by an external, localized potential, is
given by the healing length \cite{dalfovo}. Therefore,  impact of depleted atoms 
on a condensate is best given by comparison of the new term,
$V_{\rm depl}$ averaged over a few healing lengths, to  
previously dominating term in the GP equation, $aN |\phi(x)|^2$, 
averaged on the same set of points. It gives the condition (\ref{instability}).
Appearance of such a local breakdown of Bogolubov approach 
will be shown after breaking of a dark shock density profile
resulting in soliton production.

Finally, let us comment when condensate depletion might 
affect  mean-field predictions. Since  density of depleted atoms
depends on the product of $a$ and $N$ only, it is unaffected
by the limit 
$$
N\to+\infty \ \ \ , \ \ \ a\to0 \ \ \ , \ \ \ aN={\rm fixed},
$$
while the condensed part in total atomic density grows 
as 
$\langle \hat{a}_0^\dag\hat{a}_0\rangle \sim N$
(\ref{ddensity0},\ref{ddensityT}).
It means that in this limit,
being to some  extent possible with the help of Feshbach resonances
\cite{cornish},  density of depleted atoms does not affect density
measurements. The  exception might happen when there is  dynamical 
instability leading to  fast  increase of 
depleted fraction. Then sooner or later  density of
depleted atoms will start to have an important impact on condensate dynamics.

In a generic experimental situation, however, a
fraction of depleted atoms is non-negligible.
To compare a mean-field prediction to quantum many-body 
dynamics, we have chosen a particular set 
of experimentally accessible parameters -- see Appendix \ref{a1} for details.

In the following we  discuss  evolution of shock structures
resulting from white ($\eta>0$) and dark ($\eta<0$) initial density profiles,
where $\eta$ stands for relative perturbation
amplitude: as in Eq. (\ref{hyperbolicus}).
Since corrections to  mean field equations are very different 
in these two cases we discuss them separately. 

\begin{figure}
\includegraphics[angle=-0,scale=0.3, clip=true]{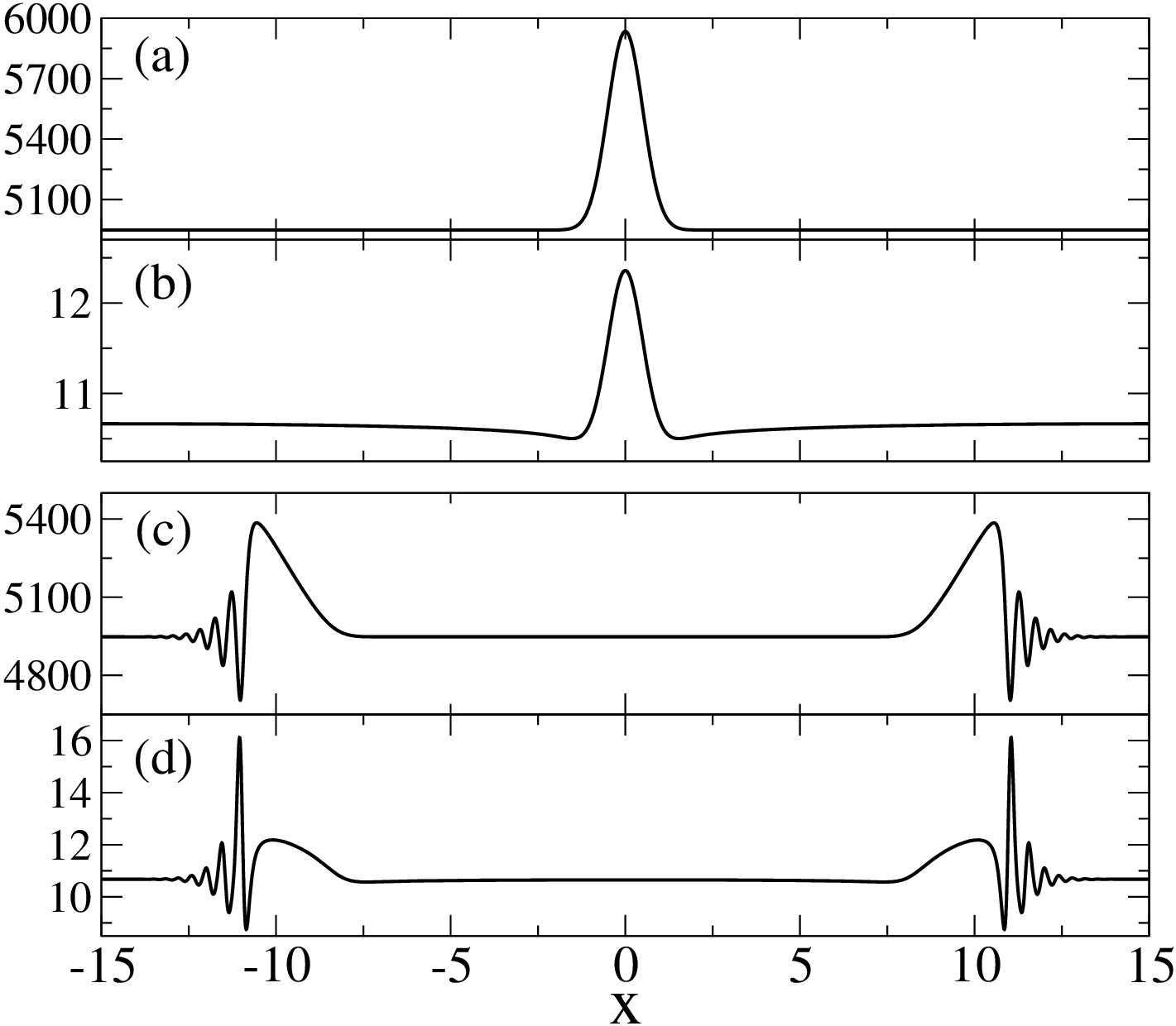}\\
\includegraphics[angle=-0,scale=0.3, clip=true]{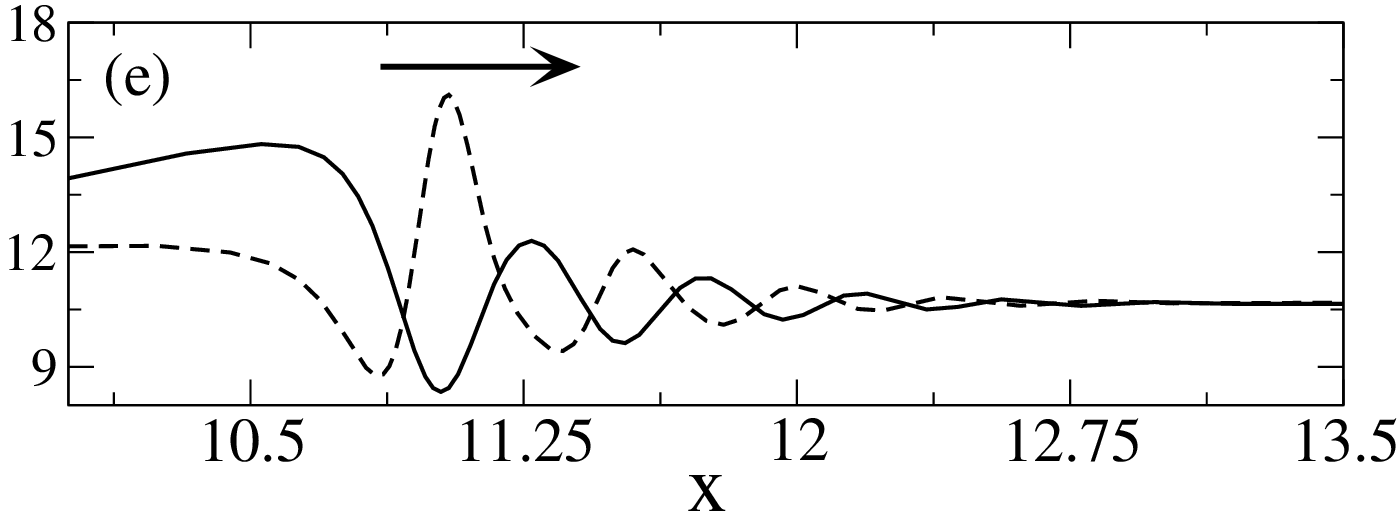}
\caption{Density of a condensate, $\langle\hat{a}_0^\dag\hat{a}_0\rangle|\phi(x)|^2$,
at $t=0, 0.6$: (a) and (c).
Density of depleted atoms ($\sum_{k=0}^{317} |v_k(x)|^2$) at $t=0, 0.6$:
(b) and (d). Plot (e): 
comparison between condensate density (solid line) and density of depleted atoms
(dashed line)
in the right-moving   shock structure;  condensate density
is rescaled for presentational purposes.  
At $t=0$ laser  potential $V_l$ was $\propto -\exp(-2x^2)$ and the system was
in a ground state;  $V_l(t>0)\equiv0$. 
The number of noncondensed atoms
equals $\sim320$ during whole evolution. Total number of
atoms is $1.5\cdot10^5$. Other parameters: $l= 15$, $aN=7500$, $T=0$.
Time of shock creation obtained from \cite{bodzio_b} is $t_s=0.36$.
For units see Appendix \ref{a1}.
}
\label{fig3}
\end{figure}

{\bf White shock waves:}
We have done different calculations changing not only the 
size of an initial perturbation but also $aN$, $N$, etc. In all 
cases we observe that  total depletion of a condensate
negligibly increases from beginning of time evolution to 
appearance of well developed shock structures. Density of depleted 
atoms around  shocks  
becomes at most ${\cal O}(1)$ larger than  depletion density
far away from perturbations.
Therefore, if density and 
distribution of depleted atoms was such that the Bogolubov approach was 
initially applicable, it will also work   during shock creation and 
propagation.

The results for experimentally relevant choice of parameters 
are depicted in Fig. \ref{fig3}. As easily seen from  differences
in scales between Figs. \ref{fig3}(a), (c) and Figs.
\ref{fig3}(b), (d), the corrections to the
mean-field result are minor. Repulsion between 
condensed and depleted atoms, results in localization of depleted atoms
around  condensate density minima: Fig. \ref{fig3}(e).
This was predicted by us in \cite{bodzio_b}, however, the amount of depleted atoms,
turns out to be insufficient to fill condensate ripples. 

We have also done 
finite temperature calculations according to (\ref{ddensityT}).
At $T\neq0$ not only quantum depletion but also thermal one
shows up, which leads to  increase of  total number of depleted
atoms in comparison to the $T=0$ situation. 
We have considered low  temperatures, i. e., $T\le 30$nK
(see Appendix \ref{a1}), 
to see whether thermal effects qualitatively affect shock 
dynamics. The only  difference with respect to $T=0$ case is
that total depletion is larger by a factor of 
${\cal O}(10)$ for $T\sim30$nK, which is still not enough for 
getting significant corrections to BEC shocks from depleted atoms.

These calculations suggest that 
the Gross-Pitaevskii equation captures  correctly physics
of white shock waves in a BEC. 
Finite temperature effects does not seem to destroy
the qualitative picture of shock formation and propagation given by the 
mean-field approach.

{\bf Dark shock waves:} Dark density profiles
undergo two distinct stages of evolution. In the first one,
shock waves form, while in the second one shock  density profile
breaks into a train of solitons that move, according to the
mean-field approach, with constant velocity and without changes of
shape. The same was independently observed numerically 
in \cite{budde} and theoretically in \cite{umarov}. 
Theoretical predictions in both  papers
are based solely on the basis of  mean-field equations. 

\begin{figure}
\includegraphics[angle=-0,scale=0.3, clip=true]{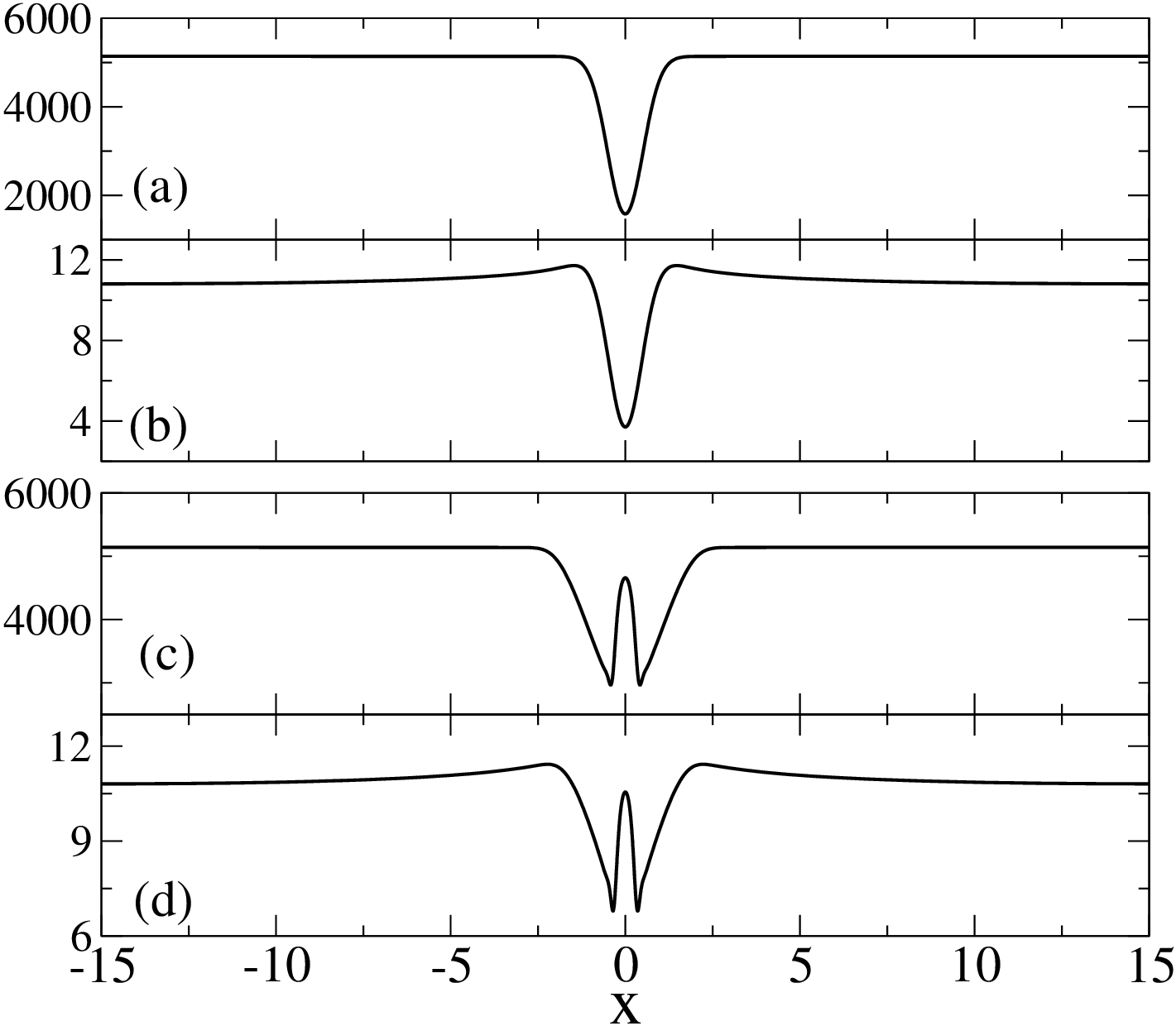}\\
\includegraphics[angle=-0,scale=0.3, clip=true]{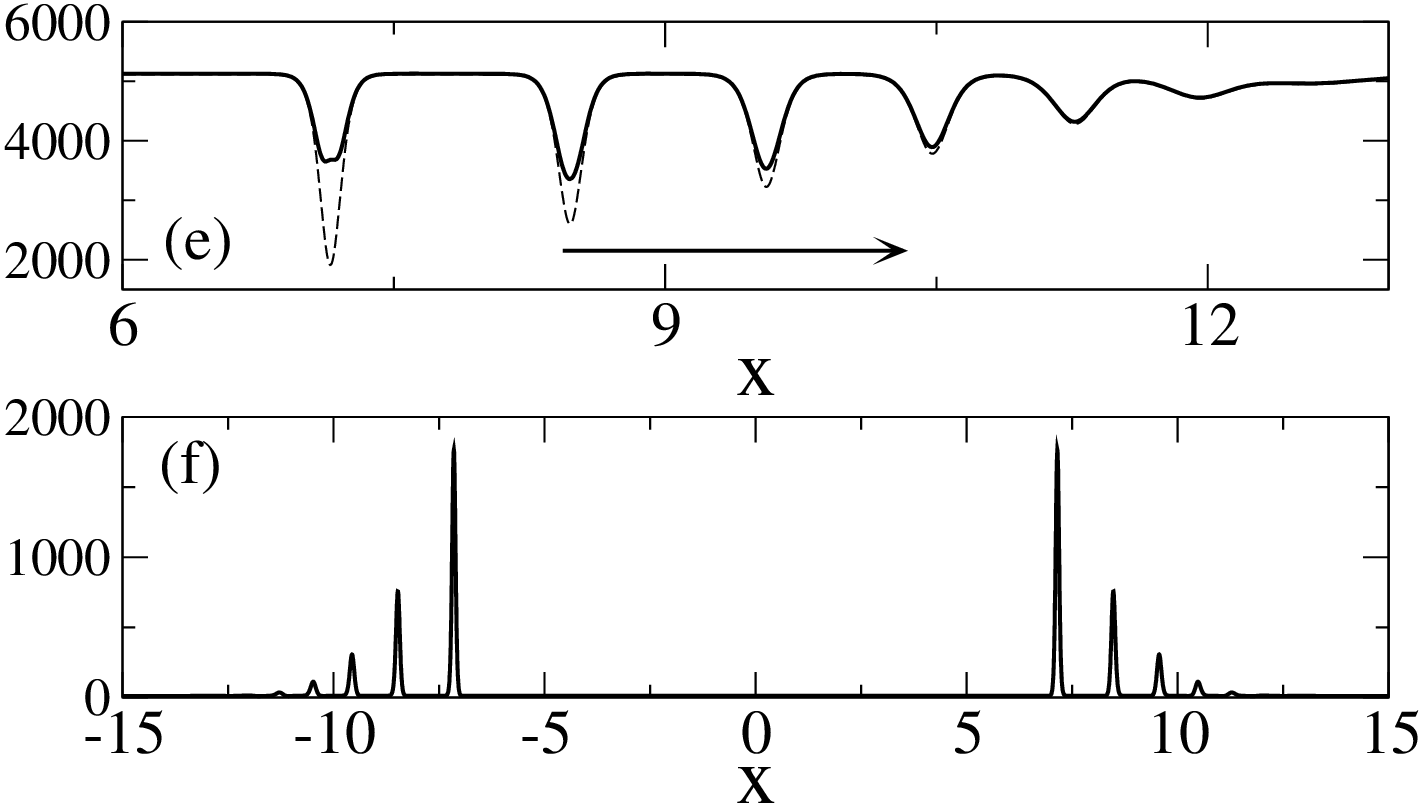}
\caption{Density of a condensate, $\langle\hat{a}_0^\dag\hat{a}_0\rangle|\phi(x)|^2$,
at $t=0, 0.06$: (a) and (c).
Density of depleted atoms
($\sum_{k=0}^{349} |v_k(x)|^2$) at $t=0, 0.06$: (b) and (d).
Plot (e): black thick line -- total density of atoms, dashed line -- condensate
density; both curves are for right-moving  dark-shock-originated
soliton train at $t=0.75$. 
Plot (f): density of depleted atoms at $t=0.75$. External laser potential
at $t=0$ was $\propto \exp(-2x^2)$ and the system was in a ground state;
$V_l(t>0)\equiv0$. 
The number of noncondensed atoms
equals $\sim320$ at $t=0$ and $\sim1000$ at $t=0.75$.
The total number of atoms equals $1.5\cdot10^5$. Other parameters: $l=15$, $aN=7500$, $T=0$.
Time of shock creation obtained from \cite{bodzio_b} is $t_s=0.086$.
For units see Appendix \ref{a1}.
}
\label{fig4}
\end{figure}

As in white shocks calculations,
we have fixed the parameters to those
experimentally relevant-- see Appendix \ref{a1}. The formation of shock 
wave structures is depicted in Figs. \ref{fig4}(a), (b), (c), (d).
Now  a rear impulse edge 
self-steepens instead of a front one. This comes from the fact that 
now impulse tail moves faster than impulse center, i. e., density minimum. 
In the case 
considered in Fig. \ref{fig4}, a shock profile breaks before 
 complete separation. Obviously, if the impulse would be more shallow
the breakdown would happen later. 
As easily seen, the  correction to  total density coming from 
depleted atoms can be neglected at this stage of time evolution.

\begin{figure}
\includegraphics[angle=-0,scale=0.3, clip=true]{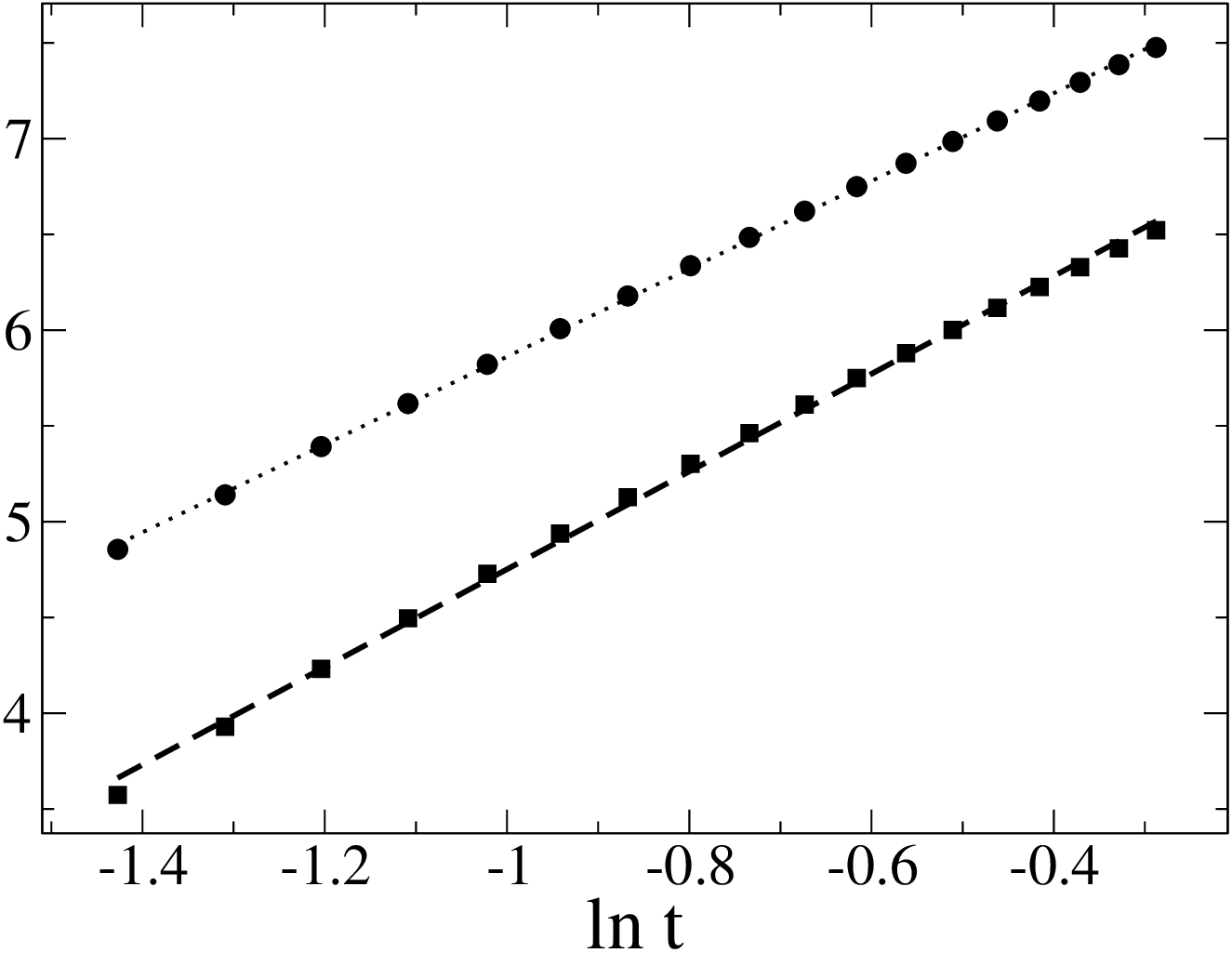}
\caption{
Introducing  notation: $p$ -
 peak density of depleted atoms in the 
deepest soliton moving to the right (Fig. \ref{fig4}e);
$dN$ - total number of depleted atoms. 
Dots: $\ln p(t)$; dotted line: a linear fit to circles, 
with slope  given by $2.29$. 
Squares: $\ln[dN(t)-dN(0)]$; dashed line: 
linear fit with  slope equal to $2.55$.
The data is plotted from $t=0.24$, i.e., a moment 
of the deepest soliton creation.
For the parameters: see Fig. \ref{fig4}, while for units see
Appendix \ref{a1}. 
}
\label{fig5}
\end{figure}

After shock breakdown,  
solitons are produced as depicted in Fig. \ref{fig4}(e). Then  
density of depleted atoms increases by orders of magnitude. It  results
in significant corrections 
to total density of atoms and causes 
local, i. e., around  soliton minima,
breakdown of the Bogolubov approach according to (\ref{instability}).
Notice that still fraction of depleted atoms, less than  $0.7\%$ in 
 Fig. \ref{fig4}(f), is so small that 
the global condition of Bogolubov approach  applicability 
is well satisfied. 
As depicted in  Fig. \ref{fig5} both  peak density of depleted atoms
at soliton minimum (dots) and  total depletion of a condensate (squares) 
follow approximately  power law  increase from the 
moment of soliton train formation. 

It is  easy to predict qualitatively what are  corrections to a
Bogolubov result coming from  depleted atoms.
The next order corrections to the Bogolubov
approach come from inclusion of repulsive  interactions between 
condensed and non-condensed atoms. It is qualitatively accounted for
by introducing a new potential term, Eq. (\ref{vdepl}), into the 
Gross-Pitaevskii equation. It means that  soliton structures
in  condensate density will become wider,  so  that depleted atoms will
have more space to distribute themselves inside  solitons. Therefore,  
increase of peak density of depleted atoms will certainly slow down. 
This observation 
is supported by nonperturbative calculation of Dziarmaga \cite{jacek}, who 
found that total density of atoms at soliton minimum approaches 
background density for large times.


Such  a fast increase in density of depleted atoms around soliton minima
is rather a generic phenomenon. Indeed, it was first theoretically
discussed  for phase-imprinted 
solitons in harmonic traps \cite{kark,solitony} as a mechanism for fast 
disappearance of solitons in the Hannover experiment \cite{hannover}.
The experimental time scale for soliton disappearance was found both 
theoretically \cite{solitony,jacek}
and experimentally \cite{hannover} to be ${\cal O}(10{\rm ms})$. Our calculation,
based on the same value of nonlinear parameter $aN=7500$,
gives the same estimation for  instant  when  corrections to dynamics 
of dark-shock-originated solitons become important. 
Indeed, the time instant $0.75$ in Fig.
\ref{fig4}(e), corresponds to $\sim12$ms-- see Appendix \ref{a1} for a unit of
time. 

To end this section we recall that predictions presented here apply to a 1D 
BEC system. There was recently an attempt to experimentally verify 
applicability of the mean-field approach to shock phenomena \cite{simula}.
The conclusion was that the ripples in a white shock wave front
are not filled with depleted atoms. Our present work 
supports this observation assuming that  1D results are
qualitatively correct in  2D rapidly rotating array of vortices on top of
which a white shock wave front propagates.

\section{Summary}
\label{sec7}

This paper presents a  complete hydrodynamical description of  basic  properties 
of shock-waves in a delta interacting 1D Bose gas. Our predictions can be 
directly verified experimentally. In principle, a proper experimental 
setup should answer at what  Lieb-Liniger parameter $\gamma$ the quantum
pressure term starts to lead to unphysical results. Notice that 
density oscillations in front of a shock wave 
are present due to this term for small enough $\gamma$,
while for large $\gamma$ they are  absent in exact solution
and shocks propagate in a very different way \cite{bodzio_f}. In fact,
a change of this kind in shock dynamics can be a nice experimental signature
of a BEC-Tonks crossover.

We also discussed
quantum many-body corrections to Gross-Pitaevskii shock-wave solutions. 
This way we clarified the role of depleted
atoms in shock dynamics. This important point was missed in previous studies of 
BEC shock waves \cite{bodzio_b,gammal,konotop,menotti,el}.

From experimental point of view,  tools for verification of our
predictions seem to be available either right now or in the nearest future. 
Indeed, a first experiment 
on sound propagation in a BEC was done  years ago \cite{ketterle}. 
Since we do not discuss time-of-flight measurements the best comparison 
between our theory and experiment should rely on {\it in situ} measurements
\cite{ketterle}. The most 
prospective experimental
setup for observation of shock dynamics in different $\gamma$ parameter ranges
is  probably provided by a box-like optical trap \cite{raizen}, where
 values of $\gamma\le1$ approaching the BEC-Tonks crossover (Fig. \ref{fig1}),
have been already achieved.
Other interesting  experimental setups include 
atom chips  \cite{jorg} and circular atom waveguides \cite{gupta}.
The latter one, being  best suited for studies of shock collisions.

From theoretical side, there are at least two possible  interesting 
extensions of this work. First of all, one can 
analyze  shock dynamics in  quasi-1D 
trapping geometries, where the system is a 3D waveguide with 
adjustable  harmonic transverse confinement. 
This can be easily done using  results from \cite{parola}.
Indeed, instead of  Eq. (\ref{muuu}) one can consider
Eq. (5) of \cite{parola} and  repeat subsequent calculations.
In the limit of tight transversal confinement, 
i. e., for a 1D Bose gas, both expressions for $\mu(\rho)$ 
lead to physically identical results.
The second interesting extension of this work 
includes  studies 
of  possible outcomes of a single shot  density measurement. In fact, 
it should be stressed that  predictions based on hydrodynamical  
and Bogolubov approaches 
apply to averages over different experimental measurements done on the system
prepared many times in the same quantum state. Needless to say,
averages may differ from  a single shot outcomes -- see \cite{juha}
for an illustrative example.

I would like to acknowledge discussions with Zbyszek
Karkuszewski.
This work was started in the Institute for Theoretical Physics
in Hannover. I'm grateful to both  
the Alexander von Humboldt Foundation for  support of this work
in Germany, and  to the US Department of Energy
for support of this research at the LANL.

\appendix
\section{Units in a box}
\label{a2}
Since this paper uses extensively results of \cite{liebliniger,lieb}
it is useful to link dimensionless quantities used by us to those of
\cite{liebliniger,lieb}. 

The  eigen equation of a 
1D Hamiltonian expressed in
terms of dimensional quantities denoted by primes is
\begin{equation}
\label{full}
-\frac{\hbar^2}{2m}\sum_i \frac{\partial^2}{\partial
{x'_i}^2}\Psi
+A' \sum_{i<j} \delta(x'_i-x'_j)\Psi=E' \Psi,
\end{equation}
to get  dimensionless Hamiltonian (\ref{manybody}) one introduces:
$x'_i=x_i l_0$, $E'= E \hbar^2/(m l_0^2)$, $A'=a \hbar^2/(m l_0)$, where
$l_0$ is an arbitrary length scale. Consideration of a time-dependent 
Schr\"odinger equation leads to $t'=t\, m l_0^2/\hbar$.

The dimensionless Hamiltonian used by Lieb and Liniger has the form
$$
\hat{H}= -\sum_i \frac{\partial^2}{\partial x_i^2}
+2 c \sum_{i<j} \delta(x_i-x_j),
$$
and can be obtained from (\ref{full}) after rescalings: $x'_i=x_i l_0$,
$E'=E \hbar^2/(2 m l_0^2)$, $A'= c \hbar^2/(m l_0)$, $t'=t \,2 m l_0^2/\hbar$.

Therefore, there are the following relations between our dimensionless quantities
and Lieb and Liniger ones marked by LL: 
$$\gamma=\gamma_{LL} \ \ , \ \ \mu=\frac{\mu_{LL}}{2} \ \ , \  \ v=\frac{v_{LL}}{2},$$
where $v$ is  sound velocity. Finally we  note, that there is a misprint 
in the
first line of  expression (1.4) providing sound velocity \cite{lieb}.  There should
be $-2\gamma de/d\gamma$ (as used in Sec. \ref{sec2})
instead of $-\gamma de/d\gamma$. 
The second line of (1.4) in \cite{lieb} is correct.

\section{A box approximation of a 3D system in the BEC limit}
\label{a1}

We aim at getting a  1D box approximation of a full 3D BEC gas confined 
in a 3D  harmonic  potential. The system in the mean-field approximation 
satisfies the following 3D Gross-Pitaevskii equation
\begin{eqnarray}
i\hbar \frac{\partial \Psi'}{\partial t'}=
&-&\frac{\hbar^2}{2m} \vec{\nabla}'^2\Psi'+ 
(m\omega^2 x'^2/2 +m\omega_{\perp}^2 r'^2/2)\Psi' \nonumber\\
&+& g'_{3D}|\Psi'|^2\Psi',
\nonumber
\end{eqnarray}
where $\omega_\perp\gg\omega$, $g'_{3D}=4\pi\hbar^2a_{sc}N/m$ with $a_{sc}$
being interatomic scattering length, and $\int d^3x'\,|\Psi'|^2 = 1$. 
First, we  rescale all
quantities using the harmonic oscillator units in the $x'$ direction.
Denoting by symbols without primes dimensionless quantities, 
$$
(x',y',z')=(x,y,z)\sqrt{\frac{\hbar}{m\omega}}, 
\Psi'=\Psi\left(\frac{m\omega}{\hbar}\right)^{3/4}, t'=\frac{t}{\omega},
$$
we arrive at the following 3D GP equation
$$
i \frac{\partial \Psi}{\partial t}= 
-\frac{1}{2} \vec{\nabla}^2\Psi+ 
(x^2/2 +\lambda^2 r^2/2)\Psi+ g_{3D}|\Psi|^2\Psi,
$$
where $\lambda$ is  trap aspect ratio $\omega_\perp/\omega$ while 
$g_{3D}= 4\pi a_{sc}N\sqrt{m\omega/\hbar}$. This 
equation  can be derived from the following energy
functional:
\begin{equation}
\label{energy}
{\cal E}[\Psi]=\int 
d^3x \left[ \frac{1}{2}
|\vec{\nabla}\Psi|^2+(x^2/2+\lambda^2r^2/2)|\Psi|^2+\frac{g_{3D}}{2}|\Psi|^4\right].
\end{equation}
To proceed further we  assume a simple form of a variational wave-function,
\begin{equation}
\label{variat}
\Psi(x,r)= \phi(x)\exp\left(-\frac{r^2}{2l_\perp^2}\right)\frac{1}{\sqrt{\pi}l_\perp},
\end{equation}
where $\int{\rm d}x\,|\phi(x)|^2=1$ and $l_\perp$ is a variational parameter.
Substituting (\ref{variat}) into (\ref{energy}) one gets
\begin{equation}
\label{energy1d}
{\cal E}[\phi]=\int dx \left[ \frac{1}{2} 
|\partial_x\phi|^2+\frac{x^2}{2}|\phi|^2+\frac{g_{3D}}{4\pi l_\perp^2}|\phi|^4\right]
+\frac{l_\perp^2\lambda^2}{2}+ \frac{1}{2l_\perp^2},
\end{equation}
which supports the following 1D GP equation 
\begin{equation}
\label{gpharm}
i \partial_t \phi=-\frac{1}{2} \partial_x^2\phi+ \frac{x^2}{2}\phi+ 
aN |\phi|^2\phi,
\end{equation}
with $aN=g_{3D}/(2\pi l_\perp^2)$.  To determine $l_\perp$, we
 substitute the Thomas-Fermi solution of 
(\ref{gpharm}), 
\begin{equation}
\label{tf}
aN|\phi(x)|^2= \frac{1}{2}\left(\frac{3aN}{2}\right)^{2/3}-\frac{x^2}{2},
\end{equation}
into (\ref{energy1d}) getting ${\cal E}(l_\perp)$, which minimization 
leads to 
$$
(g_{3D}l_\perp)^{2/3}= (250\pi^2/9)^{1/3}(\lambda^2l_\perp^4-1).
$$
In the limit of   $g_{3D}\to0$ one gets $l_\perp=1/\sqrt{\lambda}$, which is a
noninteracting cloud width. 
In a typical experiment, a cloud width is much 
larger than $1/\sqrt{\lambda}$ \cite{dalfovo},
so  
$$
l_\perp\approx\left(\frac{9}{250\pi^2}\right)^{1/10}\frac{g_{3D}^{1/5}}{\lambda^{3/5}}.
$$
For a cigar shaped cloud one can assume, e.g.,  $\omega=2\pi\cdot10$Hz, 
$\omega_\perp=2\pi\cdot569$Hz. Taking also 
$N=1.5\cdot10^5$, $a_{sc}=5.2\cdot10^{-9}$m 
(scattering length of $^{87}$Rb in the $|F=2,m_F=2\rangle$ state), 
$m= 89.91\times1.66\cdot10^{-27}$kg (atomic mass of $^{87}$Rb),   
one can find that the units of length and time are
$\sim3.4\mu m$ and $\sim16$ms, respectively. 
For these parameters $aN\approx7500$ and the unit of temperature, 
$\hbar\omega/k_B$, equals  $0.48$nK. By using all these results it  is easy to 
transform  dimensionless plots from 
Sec. \ref{sec6}, into dimensional ones.

A box approximation of dimensionally reduced harmonically trapped cloud
described above is the following. We place exactly the same number
of atoms in the box as in the quasi-1D configuration described by
(\ref{tf}). We assume that, in the absence of external laser potential,
density in a box extending from $[-l,l]$ is exactly the same as
at a  center of a harmonic trap: $[3/(4\sqrt{2})]^{2/3}/(aN)^{1/3}=1/(2l)$, which gives
$
l= (aN)^{1/3}(2/3)^{2/3}.
$

For  parameters defined above $2\cdot l\approx30$ (about $0.1$mm), which can be compared
to the Thomas-Fermi size of the harmonically trapped cloud equal here
to $\sim44$ (about $0.15$mm).
The box-like approximation leads to  results being in good qualitative 
agreement with  calculations done in a harmonically trapped case.
It concerns both  length and time scales of shock dynamics.


\begin{thebibliography}{99}

\bibitem{korepin} V.E. Korepin, N.M. Bogoliubov and A.G. Izergin,
{\it Quantum inverse scattering method and correlation functions},
(Cambridge University Press, Cambridge, 1993).

\bibitem{esslinger} H. Moritz, T. St\"oferle, M. K\"ohl and T. Esslinger,
Phys. Rev. Lett. {\bf 91}, 250402 (2003).

\bibitem{paredes} B. Paredes, A. Widera, V. Murg, O. Mandel, 
S. Folling, I. Cirac, G.V. Shlyapnikov, T.W. Hansch and I. Bloch, 
Nature {\bf 429}, 277 (2004).

\bibitem{raizen} T.P. Meyrath, F. Schreck, J.L. Hanssen,
C.-S. Chuu and M.G. Raizen,  Phys. Rev. A {\bf 71}, 041604(R) (2005).

\bibitem{bodzio_b} B. Damski, Phys. Rev. A {\bf 69}, 043610 (2004).

\bibitem{gammal} A.M. Kamchatnov, A. Gammal and R.A. Kraenkel,
Phys. Rev. A {\bf 69}, 063605 (2004). 

\bibitem{konotop} V.M. P\'erez-Garc\'ia, V. V. Konotop and V. A.
Brazhnyi, Phys. Rev. Lett. {\bf 92}, 220403 (2004).

\bibitem{menotti} C. Menotti, M. Kraemer, 
A. Smerzi, L. Pitaevskii and S. Stringari, Phys. Rev. A {\bf 70}, 023609 (2004).

\bibitem{el} G.A. El and A.M. Kamchatnov, Phys. Lett. A {\bf 350}, 192 (2006). 

\bibitem{bodzio_f} B. Damski, J.  Phys. B {\bf 37}, L85 (2004).

\bibitem{gupta} S. Gupta, K.W. Murch, K.L. Moore, T.P. Purdy and 
D.M. Stamper-Kurn, Phys. Rev. Lett. {\bf 95}, 143201 (2005).

\bibitem{olshanii}  M. Olshanii, Phys. Rev. Lett. 81, 938 (1998).

\bibitem{bfmt}  M.D. Girardeau and E.M. Wright, Laser Phys.
{\bf 12}, 8 (2002).

\bibitem{lieb} E.H. Lieb, Phys. Rev. {\bf 130}, 1616 (1963).

\bibitem{liebliniger} E.H. Lieb and W. Liniger, Phys. Rev. {\bf 130}, 1605
(1963).


\bibitem{olshani2} V. Dunjko, V. Lorent and M. Olshanii, Phys. Rev. Lett. 
{\bf 86}, 5413 (2001).

\bibitem{kohn} Y.E. Kim and A.L. Zubarev, Phys. Rev. A {\bf 67}, 015602
(2003).

\bibitem{dalfovo} F. Dalfovo, S. Giorgini, L.P. Pitaevskii and S. Stringari,
Rev. Mod. Phys. {\bf 71}, 463 (1999).

\bibitem{kolomeisky} E.B. Kolomeisky and J.P. Straley, Phys. Rev. B {\bf 46}, 
11749 (1992).

\bibitem{kolomeisky1} E.B. Kolomeisky, T.J. Newman, J.P. Straley and X. Qi, Phys. 
Rev. Lett. {\bf 85}, 1146 (2000); {\it ibid} {\bf 86}, 4709 (2001).

\bibitem{proukakis} D.J. Frantzeskakis, N.P. Proukakis, and P.G. Kevrekidis,
                    Phys. Rev. A {\bf 70}, 015601 (2004).

\bibitem{tosi} A. Minguzzi, P. Vignolo, M.L. Chiofalo, and M.P. Tosi,
               Phys. Rev. A {\bf 64}, 033605 (2001).

\bibitem{inter} M.D. Girardeau and E.M. Wright, Phys. Rev. Lett.
{\bf 84}, 5239 (2000).

\bibitem{landau} L.D. Landau and E.M. Lifshitz, {\it Fluid Mechanics}
(Pergamon, Oxford, 1989)-- see Sec. 101 for details.

\bibitem{castin} Y. Castin and R. Dum, Phys. Rev. A {\bf 57}, 3008 (1998).

\bibitem{cornish} S.L. Cornish, N.R. Claussen, J.L. Roberts, E.A.
Cornell and C.E. Wieman, Phys. Rev. Lett. {\bf 85}, 1795 (2000).


\bibitem{budde}  Z. Dutton, M. Budde, C. Slowe and L.V. Hau, Science {\bf 293},
663 (2001).

\bibitem{umarov} A.M. Kamchatnov, R.A. Kraenkel and B.A. Umarov,
Phys. Rev. E {\bf 66}, 036609 (2002).

\bibitem{jacek} J. Dziarmaga, Phys. Rev. A {\bf 70}, 063616 (2004).

\bibitem{kark} J. Dziarmaga, Z. P. Karkuszewski and K. Sacha, Phys. Rev. A
{\bf 66}, 043615 (2002).

\bibitem{solitony} J. Dziarmaga and K. Sacha, Phys. Rev. A {\bf 66}, 043620
(2002); C.K. Law, Phys. Rev. A {\bf 68}, 015602 (2003).

\bibitem{hannover} S. Burger, K. Bongs, S. Dettmer, W. Ertmer, K. Sengstock,
A. Sanpera, G.V. Shlyapnikov and M. Lewenstein, Phys. Rev. Lett. {\bf 83}, 5198 (1999).

\bibitem{simula} T.P. Simula, P. Engels, I. Coddington, V. Schweikhard, E.A.
Cornell and R.J. Ballagh, Phys. Rev. Lett. {\bf 94}, 080404 (2005).

\bibitem{ketterle} M.R. Andrews, D.M. Kurn, H.-J. Miesner, D.S. Durfee,
C.G. Townsend, S. Inouye, and W. Ketterle, Phys. Rev. Lett. {\bf 79}, 553 (1997);
{\it ibid} {\bf 80}, 2967 (1998).

\bibitem{jorg} A. Kasper, S. Schneider, C. vom Hagen, M.  Bartenstein, B.  Engeser,
T. Schumm, I. Bar-Joseph, R. Folman, L. Feenstra and J. Schmiedmayer,
J. Opt. B: Quantum Semiclass. Opt. {\bf 5}, S143 (2003).

\bibitem{parola} L. Salasnich, A. Parola and L. Reatto, Phys. Rev. A {\bf 70},
                 013606 (2004).

\bibitem{juha} J. Javanainen and S.M. Yoo, Phys. Rev. Lett. {\bf 76}, 161 (1996).

\end{thebibliography}
\end{document}